\documentclass[aps,prb,twocolumn,superscriptaddress,showpacs]{revtex4}
\usepackage{dcolumn} % Align table columns on decimal point
\usepackage{graphicx}% Include figure files
\usepackage{bm}      % bold math
\usepackage{color}
\usepackage{amsmath,amssymb}
\usepackage{hyperref}
\usepackage{multirow}
\usepackage{epstopdf}
\usepackage{latexsym}
\usepackage{subfigure}
\usepackage{wasysym} %for hexagon (in Eq environment or in text)
\usepackage{times}
\usepackage{gensymb}
\usepackage{textcomp}

\begin{document}

\title{Magnetic Properties of New Triangular Lattice Magnets A${_4}$B'B${_2}$O$_{12}$}

\author{R. Rawl}
\affiliation{Department of Physics and Astronomy, University of Tennessee, Knoxville, Tennessee 37996-1200, USA}

\author{M.~Lee}
\affiliation{Department of Physics, Florida State University, Tallahassee, Florida 32306-3016, USA}
\affiliation{National High Magnetic Field Laboratory, Florida State University, Tallahassee, Florida 32310-3706, USA}

\author{E.~S.~Choi}
\affiliation{National High Magnetic Field Laboratory, Florida State University, Tallahassee, Florida 32310-3706, USA}

\author{G.~Li}
\affiliation{School of Physics and Materials Science, Anhui University, Hefei, Anhui, 230601, People's Republic of China}

\author{K. W. Chen}
\affiliation{National High Magnetic Field Laboratory, Florida State University, Tallahassee, Florida 32310-3706, USA}

\author{R. Baumbach}
\affiliation{National High Magnetic Field Laboratory, Florida State University, Tallahassee, Florida 32310-3706, USA}

\author{C. R. dela Cruz}
\affiliation{Quantum Condensed Matter Division, Oak Ridge National Laboratory, Oak Ridge, Tennessee 37831, USA}

\author{J. Ma}
\affiliation{Department of Physics and Astronomy, Shanghai Jiao Tong University, Shanghai 200240, People's Republic of China}
\affiliation{Department of Physics and Astronomy, University of Tennessee, Knoxville, Tennessee 37996-1200, USA}

\author{H.~D.~Zhou}
\affiliation{Department of Physics and Astronomy, University of Tennessee, Knoxville, Tennessee 37996-1200, USA}
\affiliation{National High Magnetic Field Laboratory, Florida State University, Tallahassee, Florida 32310-3706, USA}

\date{\today}

\begin{abstract}
The geometrically frustrated two dimensional triangular lattice magnets A${_4}$B'B${_2}$O$_{12}$ (A = Ba, Sr, La; B' = Co, Ni, Mn; B = W, Re) have been studied by x-ray diffraction, AC and DC susceptibilities, powder neutron diffraction, and specific heat measurements. The results reveal that (i) the samples containing Co$^{2+}$ (effective spin-1/2) and Ni$^{2+}$ (spin-1) ions with small spin numbers exhibit ferromagnetic (FM) ordering while the sample containing Mn$^{2+}$ (spin-5/2) ions with a large spin number exhibits antiferromagnetic (AFM) ordering. We ascribe these spin number manipulated ground states to the competition between the AFM B'-O-O-B' and FM B'-O-B-O-B' superexchange interactions; (ii) the chemical pressure introduced into the Co containing samples through the replacement of different size ions on the A site finely tunes the FM ordering temperature of the system. We attribute this effect to the modification of the FM interaction strength induced by the change of  the O-B-O angle through chemical pressure.   
\end{abstract}
\maketitle

\begin{figure}[tbp]
	\linespread{1}
	\par
	\begin{center}
		\includegraphics[width= 3.2 in]{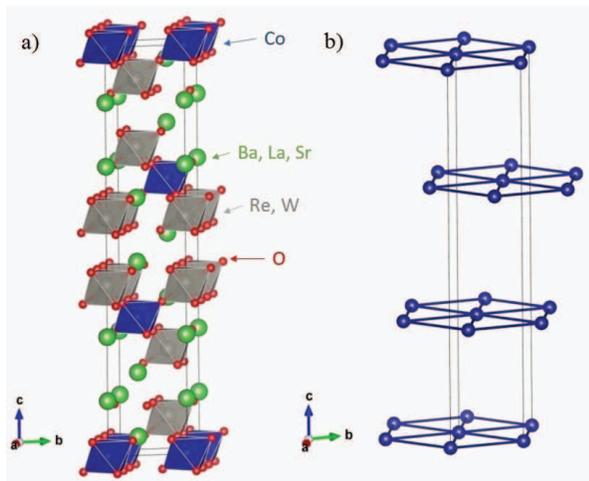}
	\end{center}
	\par
	\caption{(color online) a) Crystalline structure of A${_4}$CoB${_2}$O$_{12}$ with the space group \textit{R-3mH}. b) Staggered pattern of triangular Co$^{2+}$ planes along the $c$-axis
	}
\end{figure}

\begin{figure}[tbp]
	\linespread{1}
	\par
	\begin{center}
		\includegraphics[width= 3.2 in]{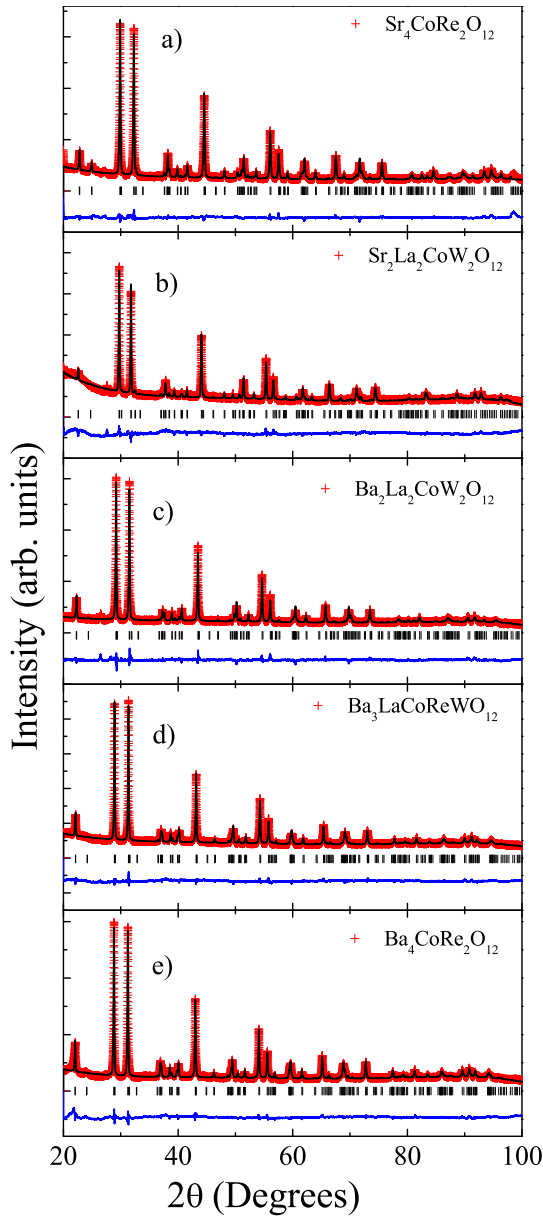}
	\end{center}
	\par
	\caption{(color online) a)-e) Powder x-ray diffraction patterns (red cross) and Rietveld refinements (black line) for five A${_4}$CoB${_2}$O$_{12}$ compounds.  The blue lines at the bottom of each panel is the difference curve. The black ticks are the reflection positions.
	}
\end{figure}

\begin{figure}[tbp]
	\linespread{1}
	\par
	\begin{center}
		\includegraphics[width= 3.2 in]{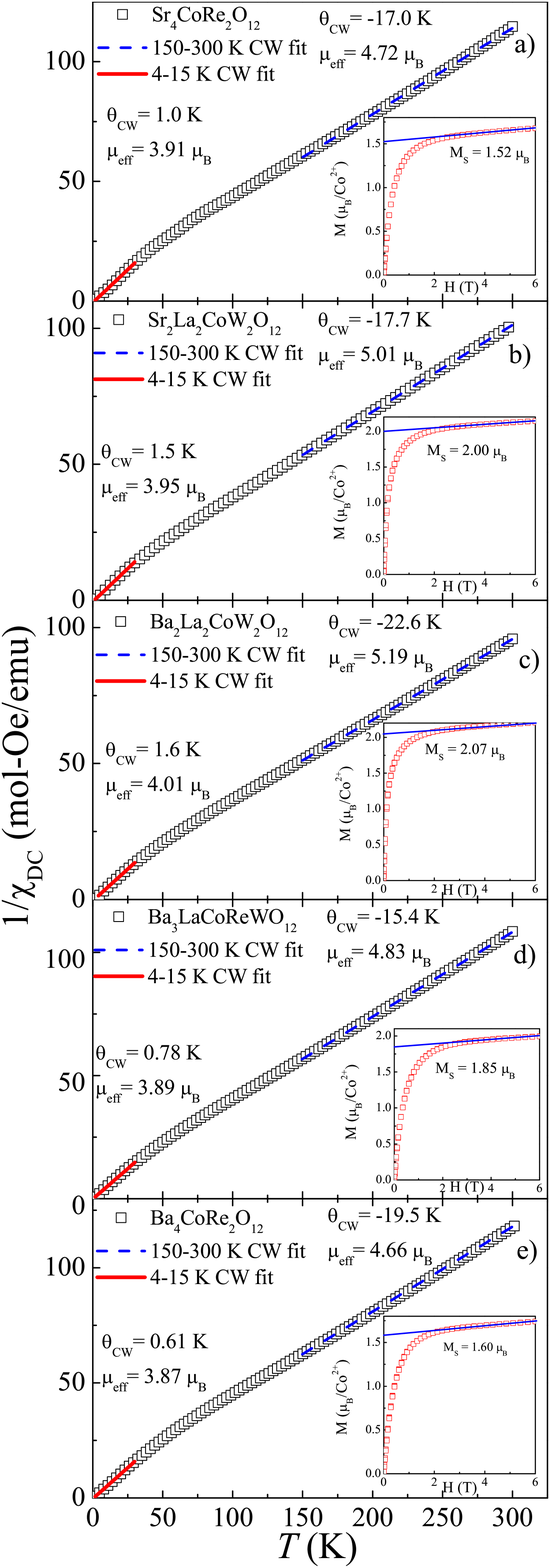}
	\end{center}
	\par
	\caption{(color online) a)-e) The inverse DC susceptibility for the A${_4}$CoB${_2}$O$_{12}$ compounds. The solid and dashed lines are Curie Weiss fittings of the low temperature and high temperature regimes, respectively. Inserts: DC magnetization taken at 1.8 K. The saturation magnetization of the Co$^{2+}$ ion is extrapolated by using a linear fit to account for the Van Vleck paramagnetic contribution.}
\end{figure}

\begin{figure}[h]
	\linespread{1}
	\par
	\begin{center}
		\includegraphics[width= 3.2 in]{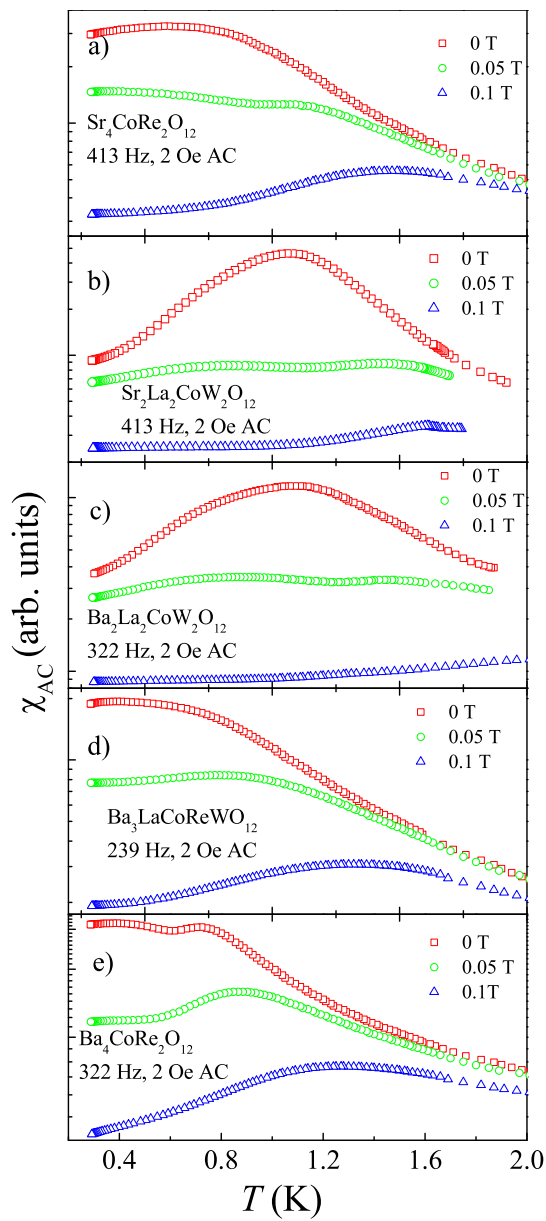}
	\end{center}
	\par
	\caption{(color online) a)-e) The real part of the AC susceptibility ($\chi_\text{AC}$) for cobalt containing A${_4}$B'B${_2}$O$_{12}$ compounds measured from 0.3 to 2.0 K under different DC magnetic fields. Excitation field of 2 Oe at low AC frequencies were used.}
\end{figure}

\begin{figure*}[tbp]
	\linespread{1}
	\par
	\begin{center}
		\includegraphics[width= 6.4 in]{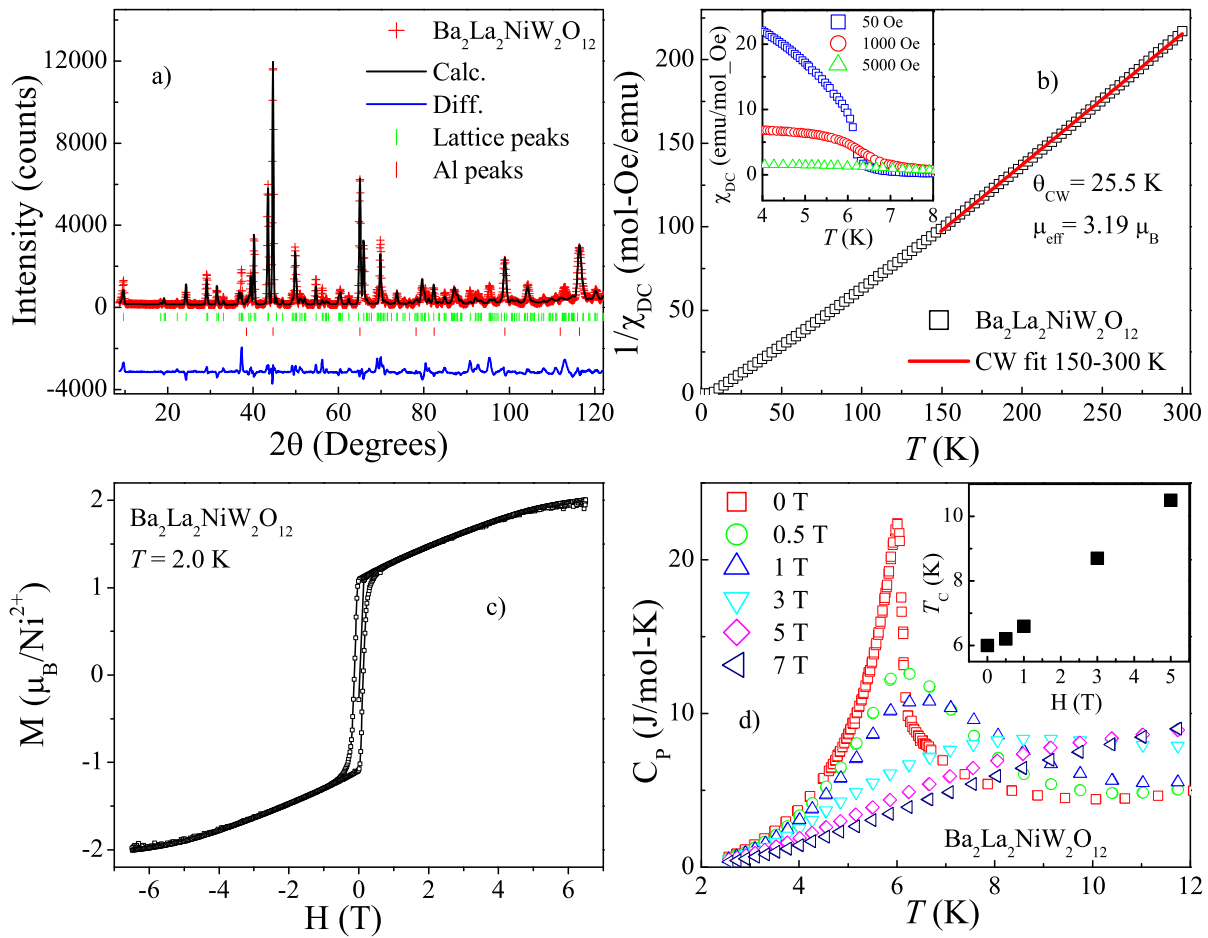}
	\end{center}
	\par
	\caption{(color online) For Ba$_2$La$_2$NiW$_2$O$_{12}$,  a) the Rietveld refinement of NPD pattern measured at room temperature using a neutron wavelength of $\lambda$ = 1.5405 \AA. b) The inverse DC susceptibility. The solid line is the linear fitting. Insert: the temperature dependence of DC susceptibility measured under different fields. c) The DC magnetization curve measured at 2 K. d) The temperature dependence of the C$_\text{P}$ measured at different DC fields. Insert: the field dependence of $T_\text{C}$.
	}
\end{figure*}

\begin{figure}[h]
	\linespread{1}
	\par
	\begin{center}
		\includegraphics[width= 3.2 in]{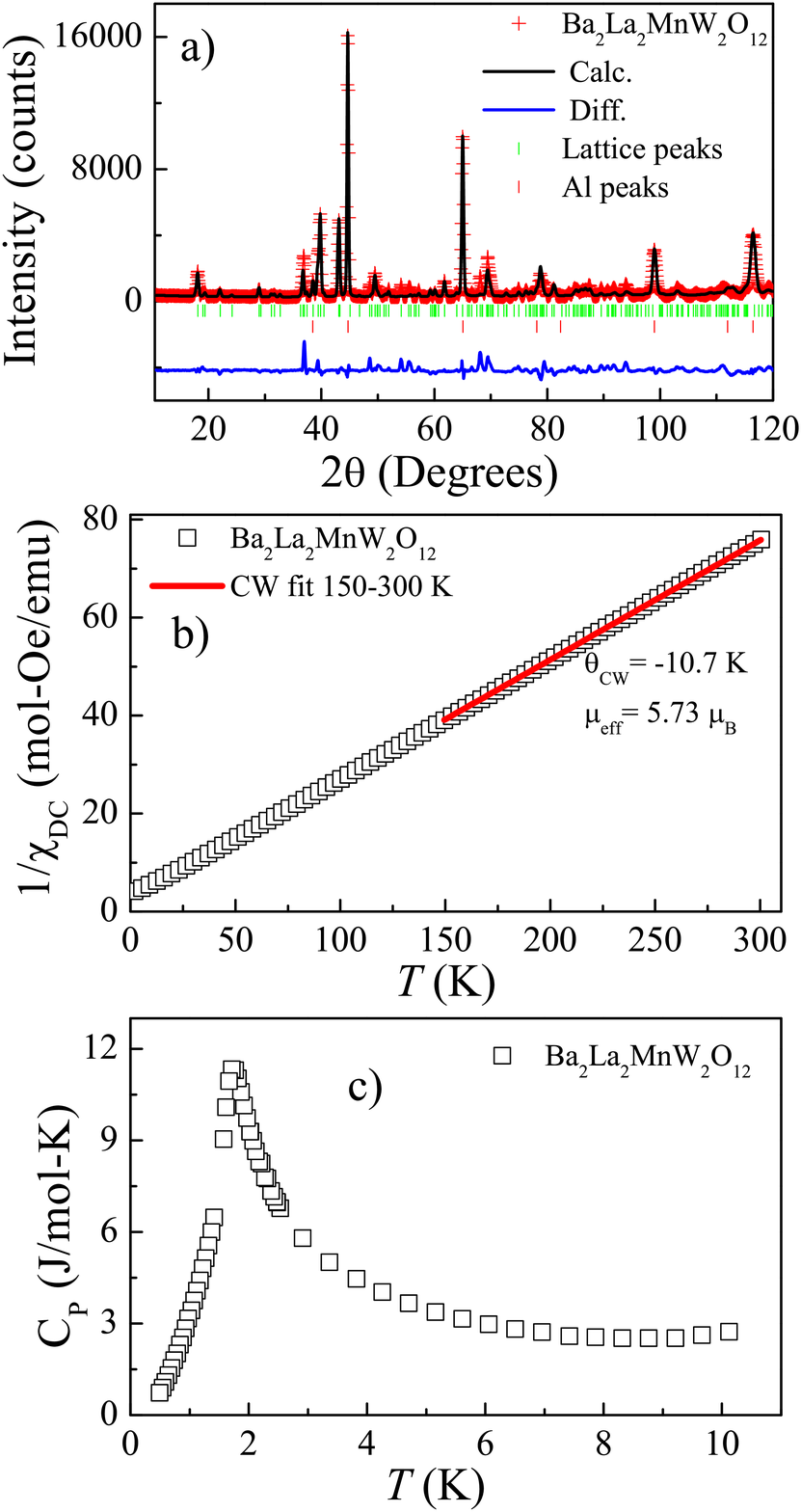}
	\end{center}
	\par
	\caption{(color online) For Ba$_2$La$_2$MnW$_2$O$_{12}$,  a) the Rietveld refinement of the NPD pattern taken at room temperature. b) The inverse DC susceptibility. The solid line is the linear fitting. c) The temperature dependence of the C$_\text{P}$ measured at zero field.
	}
\end{figure}

\begin{figure}[h]
	\linespread{1}
	\par
	\begin{center}
		\includegraphics[width= 3.2 in]{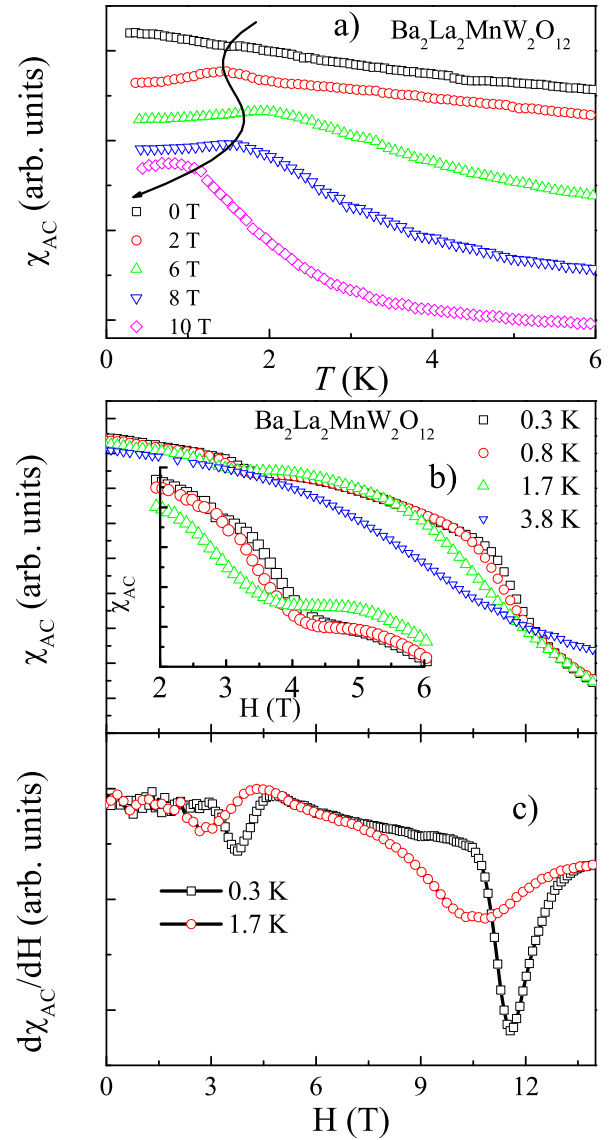}
	\end{center}
	\par
	\caption{(color online) a) The temperature dependence of $\chi_\text{AC}$ for Ba$_2$La$_2$MnW$_2$O$_{12}$ under applied DC fields.  b) The DC Field dependence of $\chi_\text{AC}$ at various temperatures. Insert: the enlargement of the data around 4 T. c) The derivative of the field dependence of $\chi_\text{AC}$ at different temperatures.
	}
\end{figure}

\begin{figure}[h]
	\linespread{1}
	\par
	\begin{center}
		\includegraphics[width= 3.2 in]{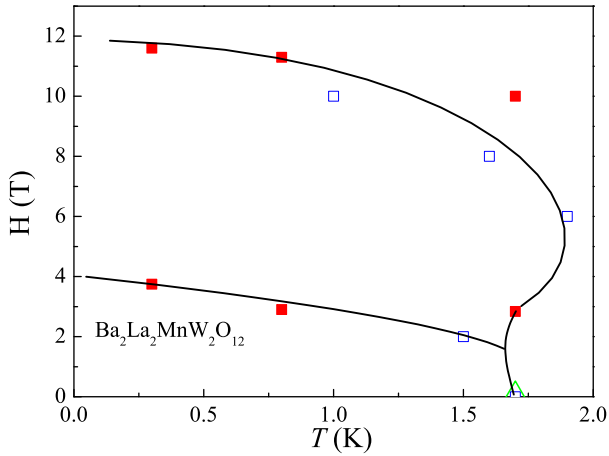}
	\end{center}
	\par
	\caption{(color online) Magnetic phase diagram of Ba$_2$La$_2$MnW$_2$O$_{12}$. Transition temperatures were found from the temperature derivative (red solid squares), the field derivative (blue open squares ) of $\chi_\text{AC}$, and zero field C$_\text{P}$ measurements (green triangle).
	}
\end{figure}

\begin{figure*}[tp]
	\linespread{1}
	\par
	\begin{center}
		\includegraphics[width= 6.8 in]{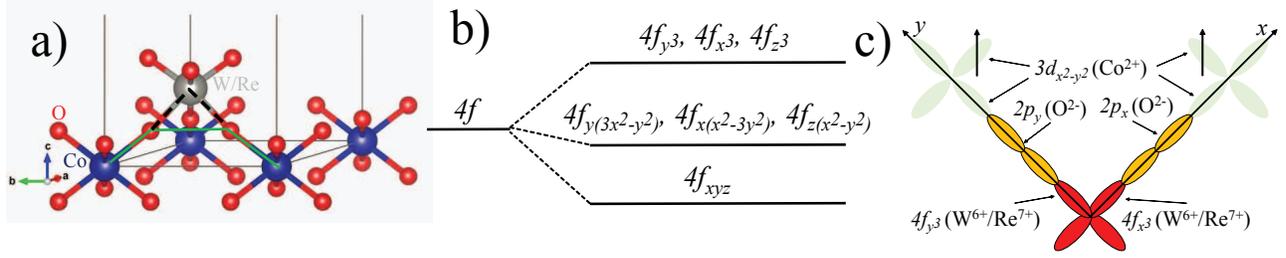}
	\end{center}
	\par
	\caption{(color online) a) Pathways for FM B'-O-B-O-B' (black dashed) and AFM B'-O-O-B' (green solid) superexchange interactions. b) Energy levels of $4f$ orbitals in cubic symmetry, with highest energy levels having the same symmetry as $p$-orbitals. c) Orbital diagram of FM B'-O-B-O-B' interaction with $f$ orbitals represented by symmetrically similar $p$ orbitals.
	}
\end{figure*}

\begin{figure}[h]
	\linespread{1}
	\par
	\begin{center}
		\includegraphics[width= 3.2 in]{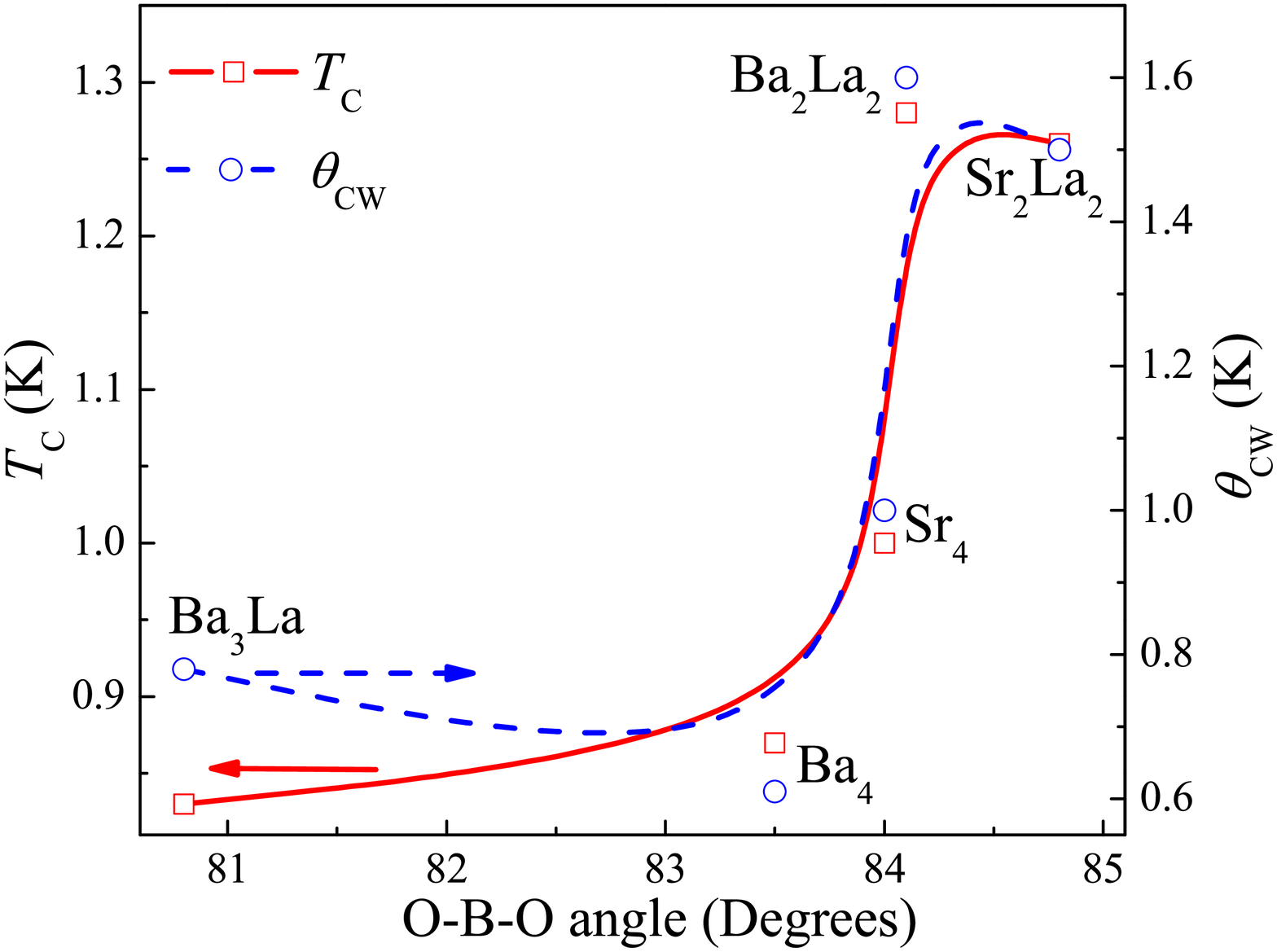}
	\end{center}
	\par
	\caption{(color online) The O-B-O bond angle dependence of  \textit{T}$_\text{C}$ and  $\theta_{\text{CW}}$.
	}
\end{figure}

\section{INTRODUCTION}

Geometrically frustrated magnets have attracted much attention for the novel properties they exhibit at low temperatures such as spin ice, spin liquids, non-collinear ground states, and fractional gauge fields \cite{ramirez,Balents,Greedan}. Of particular interest is the two dimensional (2D) triangular lattice magnets (TLMs) as, despite its simple structure, it can be host to a multitude of ground states \cite{collins,htlaf2,htlaf1}. There are several examples: (i) Ba$_3$CoSb$_2$O$_9$\cite{ba3coz,ba3cos,jeffery, nmr}, with equilateral Co$^{2+}$ (effective spin-1/2) triangular lattice, exhibits a 120 degree ordered state at zero field and an up up down phase with applied field. Although its spin wave spectrum can be reasonably described by the XXZ  model with a small easy plane anisotropy, the quantum spin fluctuations still lead to abnormal magnon decay and line broadening that cannot be accounted for by linear and nonlinear spin wave theories \cite{jma}; (ii) Ba$_3$BNb$_2$O$_9$ (B = Co\cite{2q,yokota}, Ni\cite{hwang}, Mn\cite{lee}), CuCrO$_2$\cite{cucro2,kimura}, and RbFe(MoO$_4$)$_2$\cite{RbFe,RbFe2} exhibit interesting multifeorroic properties in the 120 degree ordered state; (iii) 2H-AgNiO$_2$ \cite{AgNiO,AgNiO2} is a rare TLM that shows ordering with a collinear alternating stripe pattern which could be related to its strong easy axis anisotropy; (iv) AAg$_2$M(VO$_4$)$_2$ (A = Ba, Sr; M = Co, Ni)\cite{AgVO,BaAg} are TLMs possessing ferromagnetic (FM) orderings due to superexchange via the bridging vanadates on a triangular lattice; (v) NaVO$_2$ \cite{navo1,navo} is a rare TLM exhibiting ordering of the V$^{3+}$ orbitals. Consequently, this orbital ordering relieves the geometrical frustration and leads to long range magnetic ordering; (vi) shown recently, the exotic quantum spin liquid (QSL) state is realized in YbMgGaO$_4$ \cite{YMG1, YMG2, YMG3, YMG4} with an effective spin-1/2 Yb$^{3+}$ triangular lattice in which the spin anisotropy and next-nearest neighbor interactions play important roles.

To investigate all of these intriguing ground states, the discovery and exploration of new examples of TLMs is necessary. From a materials engineering perspective, one possible way to realize the 2D triangular lattice is by a stacked layer model, for example in the perovskite structure\cite{stack,prv}. The basic perovskite structure (ABO$_3$) can be considered as consisting of 3 layers of AO$_3$ with layers of B ions in between. The A and B site ions form corner sharing octahedra with the surrounding oxygens. Alternately, adjacent layers may consist of edge sharing oxygen octahedra, opening up further stacking mechanisms to be explored in a plethora of compounds \cite{stack}. Of interest in realizing true 2D behavior is the presence of vacant layers where neither a magnetic nor non-magnetic ion resides. This vacancy helps to ensure that interlayer interactions are small compared to intralayer interactions.

Based on this principle, we examine new 2D TLMs utilizing close-packed stacking of perovskite layers.  A promising structure is A${_4}$B\textquotesingle B${_2}$O$_{12}$ (A = Ba, Sr, La; B' = Co, Ni, Mn; B = W, Re). In this structure, using A${_4}$CoB${_2}$O$_{12}$ as the example shown in Fig. 1, the magnetic Co$^{2+}$ ions and nonmagnetic W$^{6+}$/Re$^{7+}$ ions occupy the octahedral sites in the perovskite layers in an ordered fashion at the (3$a$) and (6$c$) positions, respectively. Thus, the CoO$_6$ ocatahedral layer forms a Co$^{2+}$-triangular lattice in the $ab$ plane. Here the Co octahedra are corner sharing with the adjacent W/Re octahedra and this Co layer is sandwiched by two (W/Re)O$_6$ layers. Moreover, the octahedral interstices between the two adjacent W/Re layers are vacant to accommodate the strong electric repulsion of the W$^{6+}$/Re$^{7+}$ ions. The separation of the layers of Co$^{2+}$ by two layers of W$^{6+}$/Re$^{7+}$ and one vacant layer significantly reduces the interplane interactions and enhances the 2D nature of this structure. More intriguingly, due to the flexible chemistry in this structure, we can easily (i) introduce chemical pressure into the system by changing the lattice parameters. For example, within the set of samples, Sr$_4$CoRe$_2$O$_{12}$, Sr$_2$La$_2$CoW$_2$O$_{12}$, Ba$_2$La$_2$CoW$_2$O$_{12}$, Ba$_3$LaCoReWO$_{12}$, and Ba$_4$CoRe$_2$O$_{12}$, we expect the lattice parameters will change accordingly as we replace Sr$^{2+}$ ions with larger La$^{2+}$ and even larger Ba$^{2+}$ ions; (ii) change the spin numbers of the system by varying magnetic the B' ions. For example, within the set of samples Ba$_2$La$_2$CoW$_2$O$_{12}$, Ba$_2$La$_2$NiW$_2$O$_{12}$ and Ba$_2$La$_2$MnW$_2$O$_{12}$, the spin numbers change from effective spin-1/2 (Co$^{2+}$) to spin-1 (Ni$^{2+}$), and spin-5/2 (Mn$^{2+}$). Therefore, the A${_4}$B\textquotesingle B${_2}$O$_{12}$  structure provides an ideal platform to study the magnetic properties of new TLMs and how the perturbations, such as chemical pressure and spin numbers, affect them. So far, while the structures of several A${_4}$B\textquotesingle B${_2}$O$_{12}$  members have been reported \cite {stack2, bala1, bala2} and the susceptibility of Ba$_2$La$_2$MnW$_2$O$_{12}$\cite{balamn} has been measured down to 2 K, there are no detailed studies on their magnetic properties.

In this paper, we examine the structural and magnetic properties of these two sets of samples by using x-ray diffraction (XRD), AC and DC susceptibility ($\chi_\text{AC}$, $\chi_\text{DC}$), powder neutron diffraction, and specific heat measurements. The results reveal that the magnetic ground state is FM for samples containing Co$^{2+}$ and Ni$^{2+}$ ions with small spins but antiferromagnetic (AFM) for the Mn$^{2+}$ containing sample. We propose that this ground state change is due to the competition between the different superexchange interactions  in the structure. We also find that the chemical pressure finely tunes the transition temperature and exchange interaction in the Co samples, but does not drastically change the ground state as the spin number does.

\section{EXPERIMENTAL}

In total, seven polycrystalline samples of A${_4}$B'B${_2}$O$_{12}$ were prepared by the standard solid state reaction method. They are Sr$_4$CoRe$_2$O$_{12}$, Sr$_2$La$_2$CoW$_2$O$_{12}$, Ba$_2$La$_2$CoW$_2$O$_{12}$, Ba$_3$LaCoReWO$_{12}$, Ba$_4$CoRe$_2$O$_{12}$, Ba$_2$La$_2$NiW$_2$O$_{12}$ and Ba$_2$La$_2$MnW$_2$O$_{12}$. Stoichiometric amounts of BaCO$_3$/SrCO$_3$, La$_2$O$_3$ (pre-dried at 980 $^\circ$C overnight), CoCO$_3$/ NiO/MnO, and WO$_3$/Re metal were mixed in agate mortars, compressed into pellets, and annealed for 20 hours at temperatures of 1000 $^\circ$C for Ba$_4$CoRe$_2$O$_{12}$, 1050 $^\circ$C for Ba$_3$LaCoReWO$_{12}$, 1100 $^\circ$C for Sr$_4$CoRe$_2$O$_{12}$, and 1250 $^\circ$C for Ba$_2$La$_2$CoW$_2$O$_{12}$, Sr$_2$La$_2$CoW$_2$O$_{12}$, Ba$_2$La$_2$MnW$_2$O$_{12}$, and Ba$_2$La$_2$NiW$_2$O$_{12}$. 

Powder XRD patterns were performed at room temperature using a HUBER imaging-plate Guinier camera 670 with Ge monochromatized Cu K$\alpha$1 radiation (1.54059 \AA). High-resolution neutron powder diffraction (NPD) measurements were performed by a neutron powder diffractometer, HB2A, at the High Flux Isotope Reactor (HFIR) of the Oak Ridge National Laboratory (ORNL). Around 3 g of powder was loaded in an Al-cylinder can and mounted in a close-cycled refrigerator. We used a neutron wavelength of $\lambda$ =1.5405 $\, $\AA $\,$ with a collimation of 12$^\prime$-open-6$^\prime$. Both the XRD and NPD patterns were analyzed by the Rietveld refinement program FullProf \cite{Fprof}. DC magnetic susceptibility measurements were performed at temperatures of 2-300K using a Quantum Design superconducting interference device (SQUID) magnetometer with an applied field of 0.5 T.  AC susceptibility was measured with the conventional mutual inductance technique with a homemade setup\cite{AC}.
The specific heat data was obtained using a commercial physical property measurement system (PPMS, Quantum Design).

\section{RESULTS}

\textit{Cobalt containing compounds--} The refined XRD patterns of five Co containing compounds are shown in Fig. 2. All compounds show pure phases with the space group  \textit{R-3mH}. Used as a continuing example, Sr$_4$CoRe$_2$O$_{12}$ has lattice constants of $a$ = 5.5446(3) $\, $\AA $\,$ and $c$ = 26.7382(12) \AA.  The refined structural parameters are listed in Table I. It is obvious that the lattice parameters are governed by the size of A site ions, which increase as strontium is replaced by lanthanum and subsequently replaced by barium. We also tested the refinement by involving the site mixing between Co and W/Re sites. However, the refinements with the extra variable show no improvement from the refinementof the fully ordered structure used here. The results indicate that the site disorder in the compounds is not significant. It could be below 5\%, within our XRD resolution.

\begin{table*}[tp]
\par
\caption{Structural parameters for the Co containing compounds at room temperature (space group \textit{R-3mH}) determined from refined XRD measurements. }
\label{t1}
\setlength{\tabcolsep}{0.55cm}
\begin{tabular}{ccccccc}
\hline
\hline
Refinement & Atom & Site & {\it x} & {\it y} & {\it z} & Occupancy \\ \hline
\multirow{5}{*}{\begin{tabular}[c]{@{}c@{}}\\ Sr$_4$CoRe$_2$O$_{12}$\\ $\chi^2$ = 2.29\\ (a)\end{tabular}} & Sr1 & 6c &$   $ 0 & 0 & 0.12982(5) & 0.16666 \\
 & Sr2 & 6c &$   $ 0 & 0 & 0.29351(6) & 0.16666 \\
 & Co & 3a &$   $ 0 & 0 & 0 & 0.08333 \\
 & Re & 6c &$   $ 0 & 0 & 0.42186(3) & 0.16667 \\
 & O1 & 18h &$   $ 0.50289(50) & 0.49711(50) & 0.12312(14) & 0.50 \\
 & O2 & 18h &$   $ 0.49249(53) & 0.50751(53) & 0.29174(19) & 0.50 \\
\multicolumn{1}{l}{} & \multicolumn{1}{l}{} & \multicolumn{1}{l}{} & \multicolumn{1}{l}{} & \multicolumn{1}{l}{} & \multicolumn{1}{l}{} & \multicolumn{1}{l}{} \\
 & \multicolumn{6}{c}{$a$ = $b$ = 5.54464(26) ({\AA}), $c$ = 26.73815(126) ({\AA})} \\
\multicolumn{1}{l}{} & \multicolumn{6}{l}{} \\
\multicolumn{1}{l}{} & \multicolumn{6}{c}{Overall B-factor = 1.332 ({\AA}$^{2}$)} \\ 
\hline
\multirow{5}{*}{\begin{tabular}[c]{@{}c@{}}\\ Sr$_2$La$_2$CoW$_2$O$_{12}$\\ $\chi^2$ = 2.58\\ (b)\end{tabular}} & Sr1 & 6c &$   $ 0 & 0 & 0.13151(12) & 0.08333 \\
 & La1 & 6c &$   $ 0 & 0 & 0.13151(12) & 0.08333 \\
 & Sr2 & 6c &$   $ 0 & 0 & 0.29159(15) & 0.08333 \\
 & La2 & 6c &$   $ 0 & 0 & 0.29159(15) & 0.08333 \\
 & Co & 3a &$   $ 0 & 0 & 0 & 0.08333 \\
 & W & 6c &$   $ 0 & 0 & 0.42266(10) & 0.16667 \\
 & O1 & 18h &$   $ 0.49750(140) & 0.50250(140) & 0.12427(38) & 0.50 \\
 & O2 & 18h &$   $ 0.50466(156) & 0.49533(156) & 0.29473(48) & 0.50 \\
\multicolumn{1}{l}{} & \multicolumn{1}{l}{} & \multicolumn{1}{l}{} & \multicolumn{1}{l}{} & \multicolumn{1}{l}{} & \multicolumn{1}{l}{} & \multicolumn{1}{l}{} \\
 & \multicolumn{6}{c}{$a$ = $b$ = 5.62838(5) ({\AA}), $c$ = 26.69634(44) ({\AA})} \\
\multicolumn{1}{l}{} & \multicolumn{6}{l}{} \\
\multicolumn{1}{l}{} & \multicolumn{6}{c}{Overall B-factor = 1.127 ({\AA}$^{2}$)} \\ \hline
\multirow{5}{*}{\begin{tabular}[c]{@{}c@{}}\\ Ba$_2$La$_2$CoW$_2$O$_{12}$\\ $\chi^2$ = 2.91\\ (c)\end{tabular}} & Ba1 & 6c &$   $ 0 & 0 & 0.13426(8) & 0.08333 \\
 & La1 & 6c &$   $ 0 & 0 & 0.13425(8) & 0.08333 \\
 & Ba2 & 6c &$   $ 0 & 0 & 0.29018(9) & 0.08333 \\
 & La2 & 6c &$   $ 0 & 0 & 0.29018(9) & 0.08333 \\
 & Co & 3a &$   $ 0 & 0 & 0 & 0.08333 \\
 & W & 6c &$   $ 0 & 0 & 0.41694(8) & 0.16667 \\
 & O1 & 18h &$   $ 0.50022(89) & 0.49977(89) & 0.11969(26) & 0.50 \\
 & O2 & 18h &$   $ 0.47565(77) & 0.52436(77) & 0.29193(36) & 0.50 \\
\multicolumn{1}{l}{} & \multicolumn{1}{l}{} & \multicolumn{1}{l}{} & \multicolumn{1}{l}{} & \multicolumn{1}{l}{} & \multicolumn{1}{l}{} & \multicolumn{1}{l}{} \\
 & \multicolumn{6}{c}{$a$ = $b$ = 5.67644(22) ({\AA}), $c$ = 27.35413(106) ({\AA})} \\
\multicolumn{1}{l}{} & \multicolumn{6}{l}{} \\
\multicolumn{1}{l}{} & \multicolumn{6}{c}{Overall B-factor = 1.743 ({\AA}$^{2}$)} \\ \hline
\multirow{5}{*}{\begin{tabular}[c]{@{}c@{}}\\ Ba$_3$LaCoWReO$_{12}$\\ $\chi^2$ = 1.43\\ (d)\end{tabular}} & Ba1 & 6c &$   $ 0 & 0 & 0.12939(6) & 0.12500 \\
 & La1 & 6c &$   $ 0 & 0 & 0.12944(6) & 0.04167 \\
 & Ba2 & 6c &$   $ 0 & 0 & 0.29533(6) & 0.12500 \\
 & La2 & 6c &$   $ 0 & 0 & 0.29533(6) & 0.04167 \\
 & Co & 3a &$   $ 0 & 0 & 0 & 0.08333 \\
 & W & 6c &$   $ 0 & 0 & 0.42185(6) & 0.08333 \\
 & Re & 6c &$   $ 0 & 0 & 0.42185(6) & 0.08333 \\
 & O1 & 18h &$   $ 0.51068(82) & 0.48932(82) & 0.12143(23) & 0.50 \\
 & O2 & 18h &$   $ 0.48365(80) & 0.51635(80) & 0.29260(32) & 0.50 \\
\multicolumn{1}{l}{} & \multicolumn{1}{l}{} & \multicolumn{1}{l}{} & \multicolumn{1}{l}{} & \multicolumn{1}{l}{} & \multicolumn{1}{l}{} & \multicolumn{1}{l}{} \\
 & \multicolumn{6}{c}{$a$ = $b$ = 5.70429(9) ({\AA}), $c$ = 27.675722(54) ({\AA})} \\
\multicolumn{1}{l}{} & \multicolumn{6}{l}{} \\
\multicolumn{1}{l}{} & \multicolumn{6}{c}{Overall B-factor = 1.757 ({\AA}$^{2}$)} \\ \hline
\multirow{5}{*}{\begin{tabular}[c]{@{}c@{}}\\ Ba$_4$CoRe$_2$O$_{12}$\\ $\chi^2$ = 4.51\\ (e)\end{tabular}} & Ba1 & 6c &$   $ 0 & 0 & 0.12911(8) & 0.16666 \\
 & Ba2 & 6c &$   $ 0 & 0 & 0.29435(9) & 0.16666 \\
 & Co & 3a &$   $ 0 & 0 & 0 & 0.08333 \\
 & Re & 6c &$   $ 0 & 0 & 0.42089(8) & 0.16667 \\
 & O1 & 18h &$   $ 0.50262(108) & 0.49738(108) & 0.12325(29) & 0.50 \\
 & O2 & 18h &$   $ 0.47513(97) & 0.52487(97) & 0.28794(41) & 0.50 \\
\multicolumn{1}{l}{} & \multicolumn{1}{l}{} & \multicolumn{1}{l}{} & \multicolumn{1}{l}{} & \multicolumn{1}{l}{} & \multicolumn{1}{l}{} & \multicolumn{1}{l}{} \\
 & \multicolumn{6}{c}{$a$ = $b$ = 5.72455(33) ({\AA}), $c$ = 27.76966(161) ({\AA})} \\
\multicolumn{1}{l}{} & \multicolumn{6}{l}{} \\
\multicolumn{1}{l}{} & \multicolumn{6}{c}{Overall B-factor = 1.694 ({\AA}$^{2}$)} \\ \hline
\hline
\end{tabular}

\end{table*}

The inverse DC susceptibility for all five Co containing members is shown in Fig. 3 (a-e). All 1/$\chi_{DC}$ $\sim$ $T$ curves show a slope change around 80 K. Continuing to use Sr$_4$CoRe$_2$O$_{12}$ as an example, linear fitting from 150 K to 300 K yields a Curie-Weiss constant of $\theta_\text{CW}$ = -17 K and $\mu_\text{eff}$ = 4.72 $\mu_\text{B}$ while the linear fitting from 4.0 K to 15 K yields $\theta_\text{CW}$ = 1.0 K and $\mu_\text{eff}$ = 3.91 $\mu_\text{B}$. This change of effective magnetic moment indicates a spin state transition of the Co$^{2+}$ ions from high spin (S = 3/2) to low spin (S = 1/2). This transition is due to the octahedral environment of the Co$^{2+}$ ions, where the combination of the crystal field and spin orbital coupling lead to a Kramers doublet ground state with effective S = 1/2, as described by Low \cite{low} and further examined by Lloret et al \cite{lloret}. Thus the positive $\theta_\text{CW}$ at low temperatures suggests FM exchange interactions of the effective spin-1/2 Co$^{2+}$ ions. The spin state transition and positive  $\theta_\text{CW}$  were also observed for the four other Co compounds by similar linear fittings. The DC magnetization data was taken at 1.8 K, shown in Fig. 3 (a-e) insets. All data consistently shows a tendency of  saturation around $\mu_0$H$_s$ $\approx$  3 T. To account for Van Vleck paramagnetism, a linear fit of high field data was used to calculate the saturation magnetization (M$_\text{S}$). For Sr$_4$CoRe$_2$O$_{12}$, the extrapolation yields a value of M$_\text{S}$ = 1.52 $\mu_\text{B}$/Co$^{2+}$ and a powder-averaged gyromagnetic ratio $g$ = 3.04. For the four other compounds, the obtained $g$ is around 3 $\sim$ 4.14. All of the magnetic parameters are summarized in Table III.

The real part of the AC susceptibility measurements taken under different DC fields for the Co compounds are shown in Fig. 4. Continuing use of Sr$_4$CoRe$_2$O$_{12}$ as an example, its $\chi_{\text{AC}}$ shows a fast increase with decreasing temperature below 1.5 K followed by a broad peak. Here we define the transition temperatures as the local minima position in the first derivative of $\chi_{\text{AC}}$, which is 1.0 K for Sr$_4$CoRe$_2$O$_{12}$. With increasing DC field, this broad peak is suppressed and shifts to higher temperatures. This increase of the transition temperature under applied field suggests a FM nature which is consistent with the DC susceptibility results. The four other Co compounds show similar $\chi_{\text{AC}}$ behavior with different transition temperatures, which are listed in Table III. Therefore, all five Co compounds show FM ordering around 1 K. One noteworthy feature is that for  Sr$_2$La$_2$CoW$_2$O$_{12}$ and Ba$_2$La$_2$CoW$_2$O$_{12}$, the $\chi_{\text{AC}}$ shows a second peak at lower temperatures below 1 K when a 0.05 T DC field was applied. Whether this feature represents a second magnetic ordering to make the ordering process two-stepped or magnetic domain movements in the polycrystalline sample is not clear at this stage.

\textit{Ba$_2$La$_2$NiW$_2$O$_{12}$--} The Rietveld refinement performed on the NPD pattern (Fig. 5(a)) measured at room temperature confirms the pure phase with space group \textit{R-3mH} for Ba$_2$La$_2$NiW$_2$O$_{12}$. The refined structural parameters are $a$ = 5.6647(3) $\, $\AA $\,$ and $c$ = 27.3755(18) \AA. The detailed structural parameters are listed in Table II. The linear fitting of the inverse DC susceptibility from 150 to 300 K (Fig. 5(b)) yields a positive $\theta_{\text{CW}}$ = 25.5 K and $\mu_\text{eff}$ = 3.19 $\mu_\text{B}$, indicating FM interactions. As shown in the insert of Fig. 5(b), the $\chi_{AC}$ measured at 50 Oe clearly shows a sharp transition at  $T_\text{C}$ = 6.2 K (defined as the minimum in $d\chi_{AC}/dT$). With increasing field, this transition becomes broader and shifts to higher temperatures, which again verifies its FM nature. The DC magnetization data taken at 2.0 K (Fig. 5(c)) clearly shows a hysteresis loop and a saturation value at M$_\text{S}\approx$ 2 $\mu_\text{B}$. Under zero field, the specific heat data (C$_\text{P}$, Fig. 5(d)) shows a sharp peak at 6.2 K in agreement with $T_\text{C}$ from $\chi_{AC}$. Under applied DC fields, this peak broadens out and occurs at higher temperatures. The transition temperature (the peak position) increases linearly with the increasing field as shown in the insert of Fig. 5(d). All of this data, such as the positive $\theta_{\text{CW}}$, fast increase of $\chi_{AC}$ and peak of C$_\text{P}$ at 6.2 K, the hysteresis loop of magnetization, and the increase of  $T_\text{C}$ under fields, consistently show that Ba$_2$La$_2$NiW$_2$O$_{12}$ is a spin-1 system with long range FM ordering at 6.2 K.

\begin{table*}[tp]
\par
{\caption{Structural parameters for Ba$_2$La$_2$NiW$_2$O$_{12}$ and Ba$_2$La$_2$MnW$_2$O$_{12}$ at room temperature (space group \textit{R-3mH}) determined from refined NPD measurements. }
\label{t1}
\setlength{\tabcolsep}{0.55cm}
\begin{tabular}{ccccccc}
\hline
\hline
Refinement & Atom & Site & {\it x} & {\it y} & {\it z} & Occupancy \\ \hline
\multirow{5}{*}{\begin{tabular}[c]{@{}c@{}}\\Ba$_2$La$_2$NiW$_2$O$_{12}$\\ (a)\end{tabular}} & Ba1 & 6c &$   $ 0 & 0 & 0.13311(48) & 0.08333 \\
 & La1 & 6c &$   $ 0 & 0 & 0.13311(48) & 0.08333 \\
 & Ba2 & 6c &$   $ 0 & 0 & 0.29408(53) & 0.08333 \\
 & La2 & 6c &$   $ 0 & 0 & 0.29408(53) & 0.08333 \\
 & Ni & 3a &$   $ 0 & 0 & 0 & 0.08333 \\
 & W & 6c &$   $ 0 & 0 & 0.41928(65) & 0.16667 \\
 & O1 & 18h &$   $ 0.49819(82) & 0.50181(82) & 0.11674(28) & 0.50 \\
 & O2 & 18h &$   $ 0.47851(76) & 0.52149(76) & 0.29363(36) & 0.50 \\
\multicolumn{1}{l}{} & \multicolumn{1}{l}{} & \multicolumn{1}{l}{} & \multicolumn{1}{l}{} & \multicolumn{1}{l}{} & \multicolumn{1}{l}{} & \multicolumn{1}{l}{} \\
 & \multicolumn{6}{c}{$a$ = $b$ = 5.66471(25) ({\AA}), $c$ = 27.37551(179) ({\AA})} \\
\multicolumn{1}{l}{} & \multicolumn{6}{l}{} \\
\multicolumn{1}{l}{} & \multicolumn{6}{c}{Overall B-factor = 1.263 ({\AA}$^{2}$)} \\
\hline
Refinement & Atom & Site & {\it x} & {\it y} & {\it z} & Occupancy \\ \hline
\multirow{5}{*}{\begin{tabular}[c]{@{}c@{}}\\Ba$_2$La$_2$MnW$_2$O$_{12}$\\ (b)\end{tabular}} & Ba1 & 6c &$   $ 0 & 0 & 0.13971(71) & 0.08333 \\
 & La1 & 6c &$   $ 0 & 0 & 0.13971(71) & 0.08333 \\
 & Ba2 & 6c &$   $ 0 & 0 & 0.29399(60) & 0.08333 \\
 & La2 & 6c &$   $ 0 & 0 & 0.29399(60) & 0.08333 \\
 & Mn & 3a &$   $ 0 & 0 & 0 & 0.08333 \\
 & W & 6c &$   $ 0 & 0 & 0.41675(92) & 0.16667 \\
 & O1 & 18h &$   $ 0.48087(116) & 0.51913 (116) & 0.11743(34) & 0.50 \\
 & O2 & 18h &$   $ 0.49090(107) & 0.50910(107) & 0.29269(40) & 0.50 \\
\multicolumn{1}{l}{} & \multicolumn{1}{l}{} & \multicolumn{1}{l}{} & \multicolumn{1}{l}{} & \multicolumn{1}{l}{} & \multicolumn{1}{l}{} & \multicolumn{1}{l}{} \\
 & \multicolumn{6}{c}{$a$ = $b$ = 5.73298(37) ({\AA}), $c$ = 27.41336(275) ({\AA})} \\
\multicolumn{1}{l}{} & \multicolumn{6}{l}{} \\
\multicolumn{1}{l}{} & \multicolumn{6}{c}{Overall B-factor = 1.928 ({\AA}$^{2}$)} \\ \hline
\hline
\end{tabular}

}\end{table*}

\textit{Ba$_2$La$_2$MnW$_2$O$_{12}$--}The Rietveld refinement performed on the NPD pattern measured at room temperature (Fig. 6(a))  confirms the pure phase with space group \textit{R-3mH} for Ba$_2$La$_2$MnW$_2$O$_{12}$. The refined structural parameters are $a$ = 5.7330(4) $\, $\AA $\,$ and $c$ = 27.4134(28) \AA. The detailed structural parameters are listed in Table II. The linear fitting of the inverse DC susceptibility from 150 to 300 K (Fig. 6(c)) yields a negative $\theta_{\text{CW}}$ = -10.7 K and $\mu_\text{eff}$ = 5.73 $\mu_\text{B}$, indicating AFM interactions. The specific heat data (Fig. 6(c)) shows a sharp peak at $T_{\text{N}}$ = 1.7 K, which should represent AFM long range ordering.

The  $\chi_\text{AC}$ was measured as a function of temperature under varying applied DC fields (Fig. 7(a)) and as a function of DC field under applied temperatures (Fig. 7(b)). Here, transitions were again found using local minima of $d$$\chi_\text{{AC}}$/$dT$ and $d$$\chi_\text{AC}$/$d$H. As shown in Fig. 7(a), at zero field, the temperature dependence of $\chi_\text{AC}$  shows no significant feature but a weak slope change around 1.7 K, which is consistent with the $T_{\text{N}}$ observed from the specific heat. Then, with H = 2 T, there is a peak that appears at 1.5 K. Thereafter, this peak position shifts to higher temperatures first and then shifts to lower temperatures with increasing DC field, as indicated by the curved arrow in Fig. 7(a). Meanwhile, the field scan performed at 0.3 K shows two slope changes around 4 T and 11 T. These two features are more clearly visible as the two sharp valleys from the $d\chi_\text{AC}$/$d$H curve (Fig. 7(c)). Here we define the minima positions as H$_{\text{c1}}$ = 3.75 T and H$_{\text{c2}}$ = 11.6 T. With increasing temperature, these two features become broader and shift to lower fields. Above $T_{\text{N}}$, they disappear. A magnetic phase diagram of Ba$_2$La$_2$MnW$_2$O$_{12}$ was constructed by the transition temperatures and critical field values obtained from C$_\text{P}$ and $\chi_\text{AC}$, as shown in Fig. 8.

\section{DISCUSSION}

It is obvious that the spin number of the magnetic B' ions significantly affects the magnetic ground states of the studied A${_4}$B'B${_2}$O$_{12}$ compounds, which manifest FM ordering for the Co$^{2+}$ (S = 1/2) and Ni$^{2+}$ (S = 1) compounds but AFM ordering for the Mn$^{2+}$ (S = 5/2) compound. To understand this drastic ground state  change, we examine the superexchange interactions of the B'$^{2+}$ ions in the structure. Using the theoretical framework laid down by Kanamori\cite{Kanamori}, a qualitative description of the superexchange interaction between magnetic cations on an octahedral site can be determined from the orbital configurations of the magnetic cations and the nonmagnetic, bridging anions. In A${_4}$B'B${_2}$O$_{12}$, octahedra of B'O$_6$ are corner sharing with octahedra of BO$_6$ via an oxygen, providing two pathways for intralayer superexchange. These paths are B'$^{2+}$-O$^{2-}$-O$^{2-}$-B'$^{2+}$ with the other being B'$^{2+}$-O$^{2-}$-W$^{6+}$-O$^{2-}$-B'$^{2+}$ or B'$^{2+}$-O$^{2-}$-Re$^{7+}$-O$^{2-}$-B'$^{2+}$, as show in Fig. 9(a). The superexchange pathway along B'$^{2+}$-O$^{2-}$-O$^{2-}$-B'$^{2+}$ is commonly found in other magnetic oxides, where the interaction is AFM. 

Meanwhile, for W$^{6+}$ and Re$^{7+}$ ions on octahedral sites, the cubic crystal field splits their degenerate $4f$ orbitals (the filled outermost orbitals) into three groups as shown in Fig. 9(b). The group of $f_{x^{3}}$, $f_{y^{3}}$, and $f_{z^{3}}$ with $t_{1g}$ symmetry mainly participate in the orbital hybridization with the O$^{2-}$-$2p$ orbitals due to geometrical reasons. Thus, one possible situation for the B'$^{2+}$-O$^{2-}$-W$^{6+}$-O$^{2-}$-Cr$^{3+}$ exchange path, which is similar for the Re case, is shown in Fig. 9(c). Here we consider the superexchange interaction between the spins on the $d_{x^2-y^2}$ orbitals of the B'$^{2+}$ ions and assume that the B'$^{2+}$, O$^{2-}$ and W$^{6+}$ ions are on the same line and that the O$^{2-}$-W$^{6+}$-O$^{2-}$ bond angle is 90$^\circ$. In this situation, spin 1 on the left B'$^{2+}$ ion is transferred to the molecular orbital composed of the $p_y$ orbitals of the O$^{2-}$ 2$p$ orbitals and the $f_{y^{3}}$ orbitals of the W$^{6+}$ $4f$ orbitals; meanwhile, spin 2 on the right B'$^{2+}$ ion is transferred to the molecular orbital composed of the $p_x$ orbitals of O$^{2-}$ and the $f_{x^{3}}$ orbitals of W$^{6+}$ ions. Due to Hund's rules, these two spins on the $f_{y^{3}}$ and $f_{x^{3}}$ orbitals in the W$^{6+}$ ions have to be parallel. Then, after these two spins are transferred back to the B'$^{2+}$ ion, a FM superexchange interaction between them is formed. With these two different superexhanges in the system, it is natural to propose that the ground state change is due to the competition between them. This means the FM interaction exceeds the AFM interaction for spin-1/2 and 1 systems, but AFM interactions are stronger for the spin-5/2 system.

Since the $t_{1g}$ symmetry of the $4f$ orbitals here is identical to the symmetry of the $p$ orbitals of 4$p$ or 3$p$ orbitals, similar FM superexchange interactions should also occur for 3$d$-2$p$-4$p$(or 3 $p$)-2$p$-3$d$ paths. Several other TLMS with layered perovskite structures have been observed to possess such FM superexchange interactions. Similar competitive FM and AFM superexchanges are observed in Ba$_3$CoNb$_2$O$_9$\cite{yokota}, where the Co$^{2+}$-O$^{2-}$-Nb$^{5+}$-O$^{2-}$-Co$^{2+}$ superexchange interaction involving the Nb$^{5+}$ 4$p$ orbitals is FM, and it opposes the AFM Co$^{2+}$-O$^{2-}$-O$^{2-}$-Co$^{2+}$ interaction leading to a weaker AFM interaction. This manifests in a low AFM transition temperature as well as a low saturation field. Alternatively, AAg$_2$M(VO$_4$)$_2$ (A=Ba, Sr; M=Co, Ni)\cite{AgVO} has a stronger FM superexchange via Co$^{2+}$-O$^{2-}$-V$^{5+}$-O$^{2-}$-Co$^{2+}$ than AFM superexchange via Co$^{2+}$-O$^{2-}$-O$^{2-}$-Co$^{2+}$ and possesses an FM transition.

\begin{table*}[tbp]
\par
\caption{Summary of magnetic properties (transition temperatures \textit{T}$_\text{C}$, \textit{T}$_\text{N}$, Curie-Weiss constant $\theta_\text{CW}$, $\mu_\text{eff}$, saturation magnetization M$_\text{S}$, and gyromagnetic ratio $g$) of A${_4}$B'B${_2}$O$_{12}$ compounds }
\label{t1}
\setlength{\tabcolsep}{0.55cm}
\begin{tabular}{c|ccccc}
\hline
\hline
Compound & \textit{T}$_\text{C}$/\textit{T}$_\text{N}$ (K) & $\theta_\text{CW}$ (K) & $\mu_\text{eff}$ ($\mu_\text{B}$) & M$_\text{S}$ ($\mu_\text{B}$) & $g$ \\
\hline
Sr$_4$CoRe$_2$O$_{12}$ & 1.0 & 1.0 & 3.91 & 1.5 & 3.0\\
Sr$_2$La$_2$CoW$_2$O$_{12}$ & 1.26 & 1.5 & 3.95 & 2.0 & 4.0\\
Ba$_2$La$_2$CoW$_2$O$_{12}$ & 1.28 & 1.6 & 4.01 & 2.1 & 4.2\\
Ba$_3$LaCoReWO$_{12}$ & 0.83  & 0.78 & 3.89 & 1.9 & 3.8\\
Ba$_4$CoRe$_2$O$_{12}$ & 0.87 & 0.61 & 3.87 & 1.6 & 3.2\\
\hline
Ba$_2$La$_2$NiW$_2$O$_{12}$ & 6.2 & 25.5 & 3.19 & $\approx$ 2 & $\approx$ 2\\
\hline
Ba$_2$La$_2$MnW$_2$O$_{12}$ & 1.7 & -10.7 & 5.73 & -- & --\\
\hline
\hline
\end{tabular}
\end{table*}

Next, we look into the chemical pressure effect among the Co samples. In general, the increasing lattice parameter should decrease the exchange interactions and therefore lead to a low transition temperature. However, this is not the case here. Instead, with increasing lattice parameter, the $T_\text{C}$ first increases from Sr$_4$ to Sr$_2$La$_2$ and then to the Ba$_2$La$_2$ sample, afterwards it decreases to Ba$_3$La and to the Ba$_4$ sample. To explore the more general rule behind this chemical pressure effect, we turn to more detailed structural information. Figure 10 shows the O-W/Re-O angle dependence of $T_\text{C}$. The general trend is that with increasing angle (or as its deviation  from 90$^\circ$ decreases), the $T_\text{C}$ and $\theta_{\text{CW}}$ increases. Here, we make two assumptions: (i) $\theta_{\text{CW}}=[-zJs(s+1)]/(3k_B)$, allowing J to be extracted from $J/k_B=(2/3) \theta_{\text{CW}}$; (ii) since all Co samples show FM transitions, the dominant exchange should be FM. We assume that the chemical pressure affects the FM interaction more strongly than the AFM interaction. As we discussed above for the FM interaction, it is built on the assumption that the O$^{2-}$-W$^{6+}$-O$^{2-}$ bond angle is 90$^\circ$. The strength of this FM interaction should decrease as the O-W-O angle deviates from 90$^\circ$. The larger this deviation is, the weaker this FM interaction is, and therefore the lower $T_\text{C}$ and J should be, since we assume that the AFM interaction does not change. These are the cases for the Ba$_3$La and Ba$_4$ samples. On the other hand, the smaller this deviation is the stronger this FM interaction is and therefore the higher $T_\text{C}$ and J should be. These are the cases for the Ba$_2$La$_2$ and Sr$_2$La$_2$ samples. Therefore, this O-W/Re-O angle dependence of $T_\text{C}$ and J also confirms our proposal that the system's magnetic properties are controlled by two different superexchanges.

Finally, we comment on the phase diagram of Ba$_2$La$_2$MnW$_2$O$_{12}$. One common ground state at zero field observed for studied triangualr lattice antiferromagnets (TLAFs), such as Ba$_3$CoSb$_2$O$_9$ \cite{ba3coz} and Ba$_3$BNb$_2$O$_9$ (B = Co\cite{2q,yokota}, Ni\cite{hwang}, Mn\cite{lee}), is the 120 degree ordered state. The phase diagrams of these TLAFs consistently show a canted 120 degree, up up down, oblique, and polarized phases with increasing applied magnetic field. Comparing the phase diagram of Ba$_2$La$_2$MnW$_2$O$_{12}$ to theirs, the overall trend is similar. Therefore, we tend to ascribe its zero field ground state to be 120 degree ordering. With increasing field it enters the canted 120 degree phase below H$_{\text{c1}}$, up up down phase above H$_{\text{c1}}$, and then polarized phase above H$_{\text{c2}}$. One difference here is that we do not observe the oblique phase boundary between the up up down and polarized phases for Ba$_2$La$_2$MnW$_2$O$_{12}$. One possibility is that our AC measurements do not have the resolution to detect this phase boundary due to the polycrystalline nature of the sample.

\section{CONCLUSION}

We  have examined the structural and magnetic properties of a family of A${_4}$B'B${_2}$O$_{12}$ compounds with a 2D magnetic triangular lattice. Due to the competition between two kinds of superexchange interactions (one FM and one AFM) in the structure, the ground states of the triangular lattice can be efficiently tuned by varying the spin numbers of the magnetic B'$^{2+}$ ions, which is FM ordering for the Co and Ni samples with small spin numbers and AFM for the Mn samples with a large spin number. Moreover, the chemical pressure can finely tune the FM ordering temperature and effective exchange interactions among the Co sample by changing the local structure, mainly by changing the O-W/Re-O angle to tune the FM interaction strength. These findings demonstrate that A${_4}$B'B${_2}$O$_{12}$ is a new platform for TLMs waiting for future exploration to study geometrically frustrated magnetism.

\begin{acknowledgments}
R.R. and H.D.Z. thank the support from NSF-DMR through Award DMR-1350002. J.M. thanks the support of the Ministry of Science and Technology of China (2016YFA0300500). The research at HFIR/ORNL was sponsored by the Scientific User Facilities Division, Office of Basic Energy Sciences, US Department of Energy. The work at NHMFL is supported by NSF-DMR-1157490 and  the State of Florida.
\end{acknowledgments}


\begin{thebibliography}{50}
\bibitem{ramirez} A. P. Ramirez, Annu. Rev. Mater. Sci. {\bf 24}, 453 (1994).
\bibitem{Balents} L. Balents, Nature {\bf 464}, 199 (2010).
\bibitem{Greedan} J. Greedan, J. Mater. Chem. {\bf 11}, 37 (2000).
\bibitem{collins} M. F. Collins and O. A. Petrenko, Can. J. Phys. {\bf 75}, 605 (1997).
\bibitem{htlaf2} A. V. Chubukov and D. I. Golosov, J. Phys.:Condens. Matter {\bf 3}, 69 (1991).
\bibitem{htlaf1} O. A. Starykh, W. Jin, and A. V. Chubukov, Phys. Rev. Lett. {\bf 113}, 087204 (2014).
\bibitem{ba3coz} H. D. Zhou, C. Xu, A. M. Hallas, H. J. Silverstein, C. R. Wiebe, I. Umegaki, J. Q. Yan, T. P. Murphy, J. -H. Park, Y. Qiu, J. R. D. Copley, J. S. Gardner, and Y. Takano, Phys. Rev. Lett. {\bf 109}, 267206 (2012).
\bibitem{ba3cos} T. Susuki, N. Kurita, T. Tanaka, H. Nojiri, A. Matsuo, K. Kindo, and H. Tanaka, Phys. Rev. Lett. {\bf 110}, 267201 (2013).
\bibitem{jeffery} G. Quirion, M. Lapointe-Major, M. Poirier, J. A. Quilliam, Z. L. Dun, and H. D. Zhou, Phys. Rev. B {\bf 92}, 014414 (2015).
\bibitem{nmr} G. Koutroulakis, T. Zhou, Y. Kamiya, J. D. Thompson, H. D. Zhou, C. D. Batista, and S. E. Brown, Phys. Rev. B {\bf 91}, 024410 (2015).
\bibitem{jma} J. Ma, Y. Kamiya, T. Hong, H. B. Cao, G. Ehlers, W. Tian, C. D. Batista, Z. L. Dun, H. D. Zhou, and M. Matsuda, Phys. Rev. Lett. {\bf 116}, 087201 (2016). 
\bibitem{2q} M. Lee, J. Hwang, E. S. Choi, J. Ma, C. R. Dela Cruz, M. Zhu, X. Ke, Z. L. Dun, and H. D. Zhou, Phys. Rev. B {\bf 89}, 104420 (2014).
\bibitem{yokota} K. Yokota, N. Kurita, and H. Tanaka, Phys. Rev. B {\bf 90}, 014403 (2014).
\bibitem{hwang} J. Hwang, E. S. Choi, F. Ye, C. R. Dela Cruz, Y. Xin, H. D. Zhou, and P. Schlottmann, Phys. Rev. Lett. {\bf 109}, 257205 (2012).
\bibitem{lee} M. Lee, E. S. Choi, X. Huang, J. Ma, C. R. Dela Cruz, M. Matsuda, W. Tian, Z. L. Dun, S. Dong, and H. D. Zhou, Phys. Rev. B {\bf 90}, 224402 (2014).
\bibitem{cucro2} H. Kadowaki, H. Kikuchi, and Y. Ajiro, J. Phys.: Condens. Matter {\bf 2}, 4485 (1990).
\bibitem{kimura} K. Kimura, H. Nakamura, K. Ohgushi, and T. Kimura, Phys. Rev. B {\bf 78}, 140401 (2008).
\bibitem{RbFe} M. Kenzelmann, G. Lawes, A. B. Harris, G. Gasparovic, C. Broholm, A. P. Ramirez, G. A. Jorge, M. Jaime, S. Park, Q. Huang, A. Ya. Shapiro, and L. A. Demianets, Phys. Rev. Lett. {\bf 98}, 267205 (2007).
\bibitem{RbFe2} A. B. Harris, Phys. Rev. B {\bf 76}, 054447 (2007).
\bibitem{AgNiO} E. Wawrzy\'{n}ska, R. Coldea, E. M. Wheeler, I. I. Mazin, M. D. Johannes, T. S\"{o}rgel, M. Jansen, R. M. Ibberson, and P. G. Radaelli, Phys. Rev. Lett. {\bf 99}, 157204 (2007).
\bibitem{AgNiO2} E. M. Wheeler, R. Coldea, E. Wawrzy\'{n}ska, T. S\"{o}rgel, M. Jansen, M. M. Koza, J. Taylor, P. Ardroguer, and N. Shannon, Phys. Rev. B {\bf 79}, 104421 (2009).
\bibitem{AgVO} A. M\"{o}ller, N. E. Amuneke, P. Daniel, B. Lorenz, C. R. Dela Cruz, M. Gooch, and P. C. W. Chu, Phys. Rev. B {\bf 85}, 214422 (2012).
\bibitem{BaAg} A. A. Tsirlin, A. M\"{o}ller, B. Lorenz, Y. Skourski, and H. Rosner, Phys. Rev. B {\bf 85}, 014401 (2012).
\bibitem{navo1} T. M. McQueen, P. W. Stephens, Q. Huang, T. Klimczuk, F. Ronning, and R. J. Cava, Phys. Rev. Lett. {\bf 101}, 166402 (2008).
\bibitem{navo} T. Jia, G. R. Zhang, Z. Zeng, and H. Q. Lin, Phys. Rev. B {\bf 80}, 045103 (2009).
\bibitem{YMG1} Y. Li, G. Chen, W. Tong, L. Pi, J. Liu, Z. Yang, X. Wang, and Q. Zhang, Phys. Rev. Lett. {\bf 115}, 167203 (2015).
\bibitem{YMG2} Y. Li, H. Liao, Z. Zhang, S. Li, F. Jin, L. Ling, L. Zhang, Y. Zou, L. Pi, Z. Yang, J. Wang, Z. Wu, and Q. Zhang, Scientific Reports {\bf 5}, 16419 (2015).
\bibitem{YMG3} Y. Shen, Y. D. Li, H. Wo, Y. Li, S. Shen, B. Pan, Q. Wang, H. C. Walker, P. Steffens, M. Boehm, Y. Hao, D. L. Quintero-Castro, L. W. Harriger, M. D. Frontzek, L. Hao, S. Meng, Q. Zhang, G. Chen, and J. Zhao, Nature {\bf 540}, 559 (2016).
\bibitem{YMG4} J. A. M. Paddison, M. Daum, Z. Dun, G. Ehlers, Y. Liu, M. B. Stone, H. D. Zhou, and M. Mourigal, Nature Physics {\bf 13}, 117 (2016). 
\bibitem{stack} L. Katz and R. Ward, Inorg. Chem. {\bf 3}, 205 (1964).
\bibitem{prv} A. Maignan, W. Kobayashi, S. H\'{e}bert, G. Martinet, D. Pelloquin, N. Bellido, and Ch. Simon, Inorg. Chem. {\bf 47}, 8553 (2008).
\bibitem{stack2} J. M. Longo, L. Katz, and R. Ward, Inorg. Chem. {\bf 4}, 235 (1965).
\bibitem{bala1} H. J. Rother, A. Fadini, and S. Kemmler-Sack, Z. Anorg. Allg. Chem {\bf 463}, 137 (1980).
\bibitem{bala2} S. Kemmler-Sack, Z. Anorg. Allg. Chem {\bf 461}, 142 (1980).
\bibitem{balamn} Z. F. Li, J. L. Sun, L. P. You, Y. X. Wang, and J. H. Lin, J. Alloy. Compd. {\bf 379}, 117 (2004).
\bibitem{Fprof} J. Rodriguez-Carvajal, Physica B {\bf 192}, 55 (1993).
\bibitem{AC} Z. L. Dun, M. Lee, E. S. Choi, A. M. Hallas, C. R. Wiebe, J. S. Gardner, E. Arrighi, R. S. Freitas, A. M. Arevalo-Lopez, J. P. Attfield, H. D. Zhou, and J. G. Cheng, Phys. Rev. B {\bf 89}, 064401 (2014).
\bibitem{low} W. Low, Phys. Rev. {\bf 109}, 256 (1958).
\bibitem{lloret} F. Lloret, M. Julve, J. Cano, R. Ruiz-Garc\'{i}a, and E. Pardo, Inorg. Chim. Acta. {\bf 361}, 3432 (2008).
\bibitem{Kanamori} J. Kanamori, J. Phys. Chem. Solids {\bf 10}, 87 (1959).
\end{thebibliography}
\end{document}